\documentclass[12pt,notitlepage]{article}
\usepackage[dvips]{graphicx}
\usepackage{titling}

\begin{document}

\long \def \blockcomment #1\endcomment{}

\date{March 2020}

\title{Four Metrics\thanks{ Essay written for the Gravity
  Research Foundation
  2020 Awards for Essays on Gravitation}}

\author{Michael Creutz\\
  Senior Physicist Emeritus\\
Physics Department\\
  Brookhaven National Laboratory\\
  Upton, NY 11973, USA\\
\\
  {mike@latticeguy.net}\\
}  

\begin{titlingpage}
\maketitle

\begin{abstract}
  { A central idea in general relativity is that physics should not
    depend on the space-time coordinates in use \cite{einstein}.  But
    the qualitative description of various phenomena can appear
    superficially quite different.  Here we consider falling into a
    classical black hole using four distinct but equivalent metrics.
    First is the Schwarzchild case, with extreme time dilation at the
    horizon.  Second, rescaling the dilation allows falling into the
    hole in finite proper time.  Third, time and space are rescaled
    into a Penrose motivated picture where light trajectories all have
    unit slope.  Fourth, a white hole variation of the second metric
    allows passage out through the horizon, with reentry forbidden.  }
\end{abstract}
\end{titlingpage}


\blockcomment

The equations of general relativity are conventionally written in a
form that is invariant under choice of coordinates \cite{einstein}.
For an extreme example, one might use either Copernican or Ptolemaic
coordinates to describe the solar system. It is just that the details
for some phenomena might appear simpler with one or the other.  Here
we use four different metrics to describe an object falling into a
black hole.  In all cases the physics is equivalent, but the
qualitative appearance is quite different.

First is the standard Schwarzchild metric.  This displays extreme time
dilation near the horizon.  Second as a smoothed metric, where a
falling object does not see a singularity at the horizon, but in its
own proper time proceeds smoothly into the black hole interior.  Third
we use the smoothed metric as a route to Penrose like coordinates,
where radial light trajectories are all either parallel or orthogonal.
Finally the fourth set of coordinates describes what is sometimes
called a ``white hole.'' Here physical trajectories can emerge from
the horizon, however in the other coordinate systems this occurs in
the infinite past.


\endcomment

\section{The Schwarzchild metric}

We start with the famous Schwarzchild \cite{schwarzchild} formula for
space-time around a stationary and non-rotating gravitating object
\begin{equation}
  ds^2=(1-1/r) dt^2  -(1-1/r)^{-1} dr^2
  -r^2(d\theta^2+\sin^2(\theta)d\phi^2).
\label{metric}
\end{equation}
Here we use polar coordinates for the spatial components
\begin{eqnarray*}
  x=r\sin(\theta)\cos(\phi),\qquad
  y=r\sin(\theta)\sin(\phi),\qquad
  z=r\cos(\theta).
  \end{eqnarray*}
For simplicity, measure distances in units of the ``Schwarzchild
radius,'' {\em i.e.} the radius of the horizon
is taken as unity.

Light trajectories are defined by $ds^2=0$.  For radial motion we have
\begin{equation}
  {dt \over dx}=\pm (1-1/r)^{-1}.
\end{equation}
The plus (minus) sign is for an out-going (in-going) wave.  This integrates to
\begin{equation}
  t-t_0=\pm\left(r-r_0+\log\left({r-1 \over r_0-1}\right)\right).
  \end{equation}
Here $r_0$ and $t_0$ locate the initial point for the trajectory.

\begin{figure}
  \centering
  \includegraphics[width=.9 \hsize]{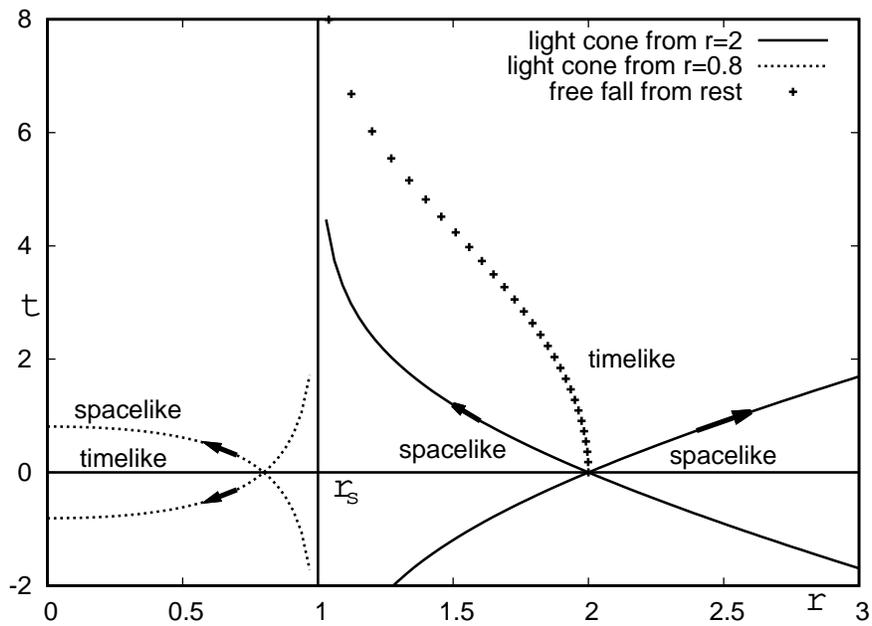}
  \caption{Light cones in Schwarzchild coordinates in the vicinity of
    a black hole.  The radius of the horizon is taken as unity.  One
    cone shown starts outside the hole at radius 2.  The second starts
    at radius 0.8, inside the black hole.  Note how the cone inside
    the hole is turned sideways relative to the outer one.  The figure
    also shows the trajectory of a freely falling object started at
    $r=2$, with tick marks indicating constant intervals in its proper
    time.}
  \label{slight}
  \end{figure}

Remarkably as $r$ approaches unity for an ingoing wave, time goes to
infinity!  The light never reaches the horizon.  With this metric,
neither light can leave the black hole, nor can it enter it.  As
nothing goes faster than light, nothing else can reach the horizon in
finite Schwarzchild time.  This behavior is sketched in
Fig. \ref{slight}.  The figure also shows the path of a freely falling
object initially at rest at $\{t,r\}=\{0,2\}.$ With the falling
trajectory, tick marks are indicated at intervals of $0.1$ in proper
time $s$ as measured on the corresponding path.  Note how these marks
spread as the time variable $t$ increases.  Time dilates as the black
hole is approached.  The spread increases rapidly with time, and
$t=\infty$ is reached at finite proper time.

\vfill\eject

\section{Falling through the horizon}

As is well known, the singularity in the Schwarzchild metric at the
horizon is somewhat artificial. As first made clear by Lemaitre
\cite{lemaitre}, we can smooth the singularity at the horizon with an
appropriate redefinition of coordinates.

\begin{figure}
  \centering \includegraphics[width=.9 \hsize]{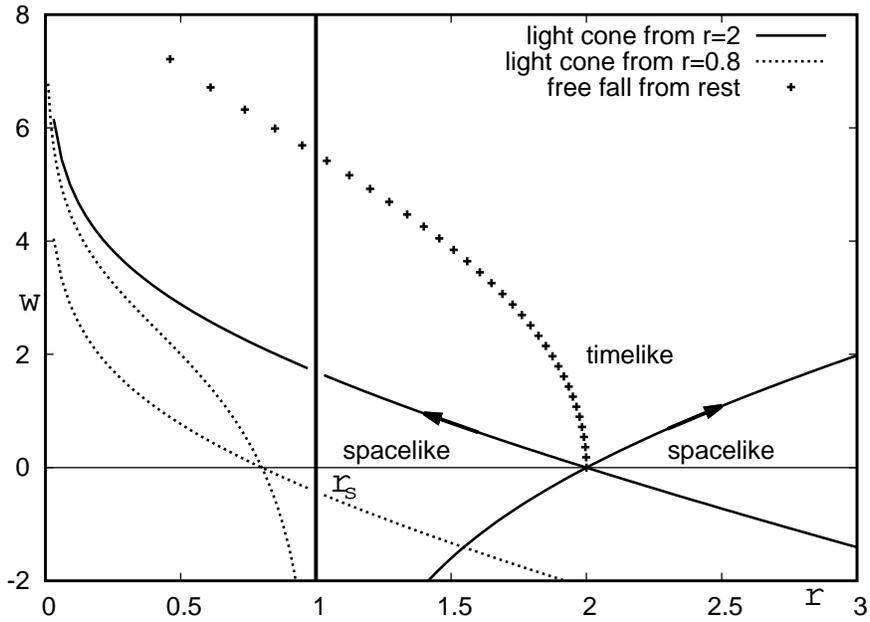}
  \caption{Light cones in the modified coordinates $w,r$ in the
    vicinity of a black hole.  As before, the Schwarzchild radius is
    taken as unity.  One cone shown starts outside the hole at radius
    2.  The second starts at radius 0.8, inside the black hole.  Note
    how the incoming light wave smoothly crosses the horizon but never
    reaches the origin.  As in Fig. \ref{slight}, the path of a freely
    falling object released at $r=2$ is also shown. }
  \label{wlight}
\end{figure}

For a simple example, consider a new definition of time, call it $w$,
obtained by a change of variables similar to that made by Finkelstein
\cite{finkelstein}
\begin{equation}
  w=t+\log(1-1/r).
\end{equation}
The addition is singular at the horizon, $r=1$, but this serves to
remove the unphysical singularity in the Schwarzchild metric.  The
metric equation now becomes
\begin{equation}
  ds^2=dw^2\ ( 1-1/r)-2dw\ dr /r^2
-dr^2\ (1+1/r)(1+1/r^2)
-r^2(d\theta^2+\sin^2(\theta)d\phi^2).
\end{equation}
This is smooth at the horizon.  The coordinate $w$ can properly be
considered as an alternate ``time'' in the sense that constant $w$
surfaces continue to have negative $ds^2$; {\it i.e.} they remain
space-like.  The horizon is still well defined as a ``separatrix,''
dividing ``out-going'' light-waves into those that ultimately head
towards $r=0$ or $r=\infty$.  This behavior is sketched in
Fig. \ref{wlight}.

This figure is not symmetric when inverting the time coordinate $w$.
Any time translation invariant coordinate system allowing crossing of
the horizon is necessarily not time reversal symmetric.  The time
reversed black hole is what is sometimes called a ``white hole,'' to
which we return later.

As with the Schwarzchild coordinates, this figure also shows the
trajectory for a freely falling mass starting at radius 2.  Again,
this is marked with tick marks at constant separation in proper time.
Now the gravitational red shift continues to increase even inside the
hole, with the object reaching $w=\infty$ in a finite proper time.


\vfill\eject

\section{Penrose coordinates}

Penrose \cite{Penrose:1964ge} suggested modifying coordinates as an
aide to visualizing black hole geometry.  Working in polar coordinates
to eliminate angular degrees of freedom, it is possible to distort the
coordinates so that radial light rays always run along parallel
straight lines, with inward and outward null curves remaining
perpendicular.  Once such coordinates are established, distances are
rescaled to map space and time into a finite range.

\begin{figure}
  \centering
  \includegraphics[width= \hsize]{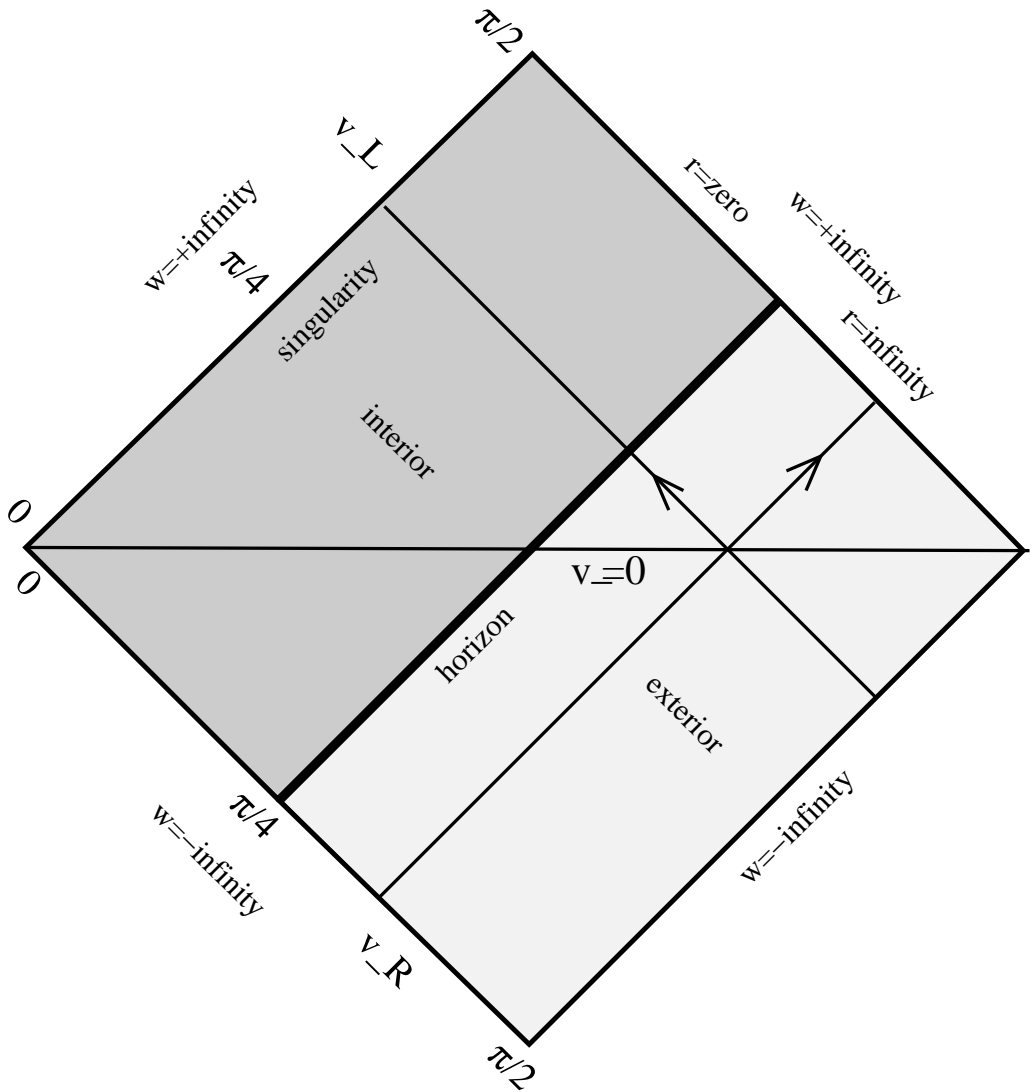}
  \caption{Redrawing the area around the black hole in the new
    coordinates $v_\pm$.  In these coordinates the horizon becomes a
    straight diagonal line.  Out-going null trajectories are all
    parallel to the horizon while in-going ones always cross the
    horizon.}
  \label{penrose}
\end{figure}

The resulting diagram is not unique, however such a construction is
particularly simple using the coordinates $\{w,r\}$ of the previous
section.  Starting with any given point in this plane, construct
in-going and out-going light-like curves.  All such curves can be
obtained by translating in $w$ the lines in Fig. \ref{wlight}, and all
cross the $w=0$ axis somewhere.  Refer to the crossing points as
$r_{R},$ and $r_L$ for the outgoing and ingoing light-like curves,
respectively.  (Inside the horizon the ``out-going'' light actually
moves inward, but more slowly than the ``in-going'' one.)  The
mapping between $\{w,r\}$ and $\{r_L,r_R\}$ is one to one, and we can
use the latter as intermediate coordinates to describe the full
space-time.  Positive (negative) $w$ corresponds to $r_L>r_R$
\ ($r_L<r_R$).

In these coordinates out-going light waves follow constant $r_R$ and
in-movers have constant $r_L$.  Now perform a scaling to map infinity
to a finite value.  For this define
\begin{equation}
v_{L,R}=\arctan(r_{L,R}).
  \end{equation}
Since the coordinates $\{r_L,r_R\}$ each range from 0 to $\infty$, the
new variables satisfy $0<v_{L,R}<\pi/2$.  Finally, it is conventional
to rotate the resulting diagram by $\pi/4$ by defining
\begin{equation}
v_\pm=v_R\pm v_L
\end{equation}
where $v_-$ now represents time and $v_+$ space.  This gives the
resulting picture in Fig.~\ref{penrose}.

Now in-going light rays are at an angle of $3\pi/4$ and always cross
the horizon.  Out-going waves are at $\pi/4$ and never touch the
horizon.  the singularity at $r=0$ is mapped onto the top of the
diagram, which any in-going falling object asymptotically approaches.


\vfill\eject

\section{Emerging from a white hole}
The metric using $w$ for time is not symmetric under reversal of the
sign of $w$.  But reversing this time is effectively a definition of
yet another time, $\tau$, defined from the Schwarzchild time by
\begin{equation}
  \tau=t-\log(1-1/r).
\end{equation}
for which the metric becomes
\begin{equation} 
  ds^2=d\tau^2\ ( 1-1/r)+2dz\ dr /r^2
  -dr^2\ (1+1/r)(1+1/r^2)
-r^2(d\theta^2+\sin^2(\theta)d\phi^2).
\end{equation}
Again, constant $\tau$ surfaces are space-like.

\begin{figure}
  \centering \includegraphics[width=.9 \hsize]{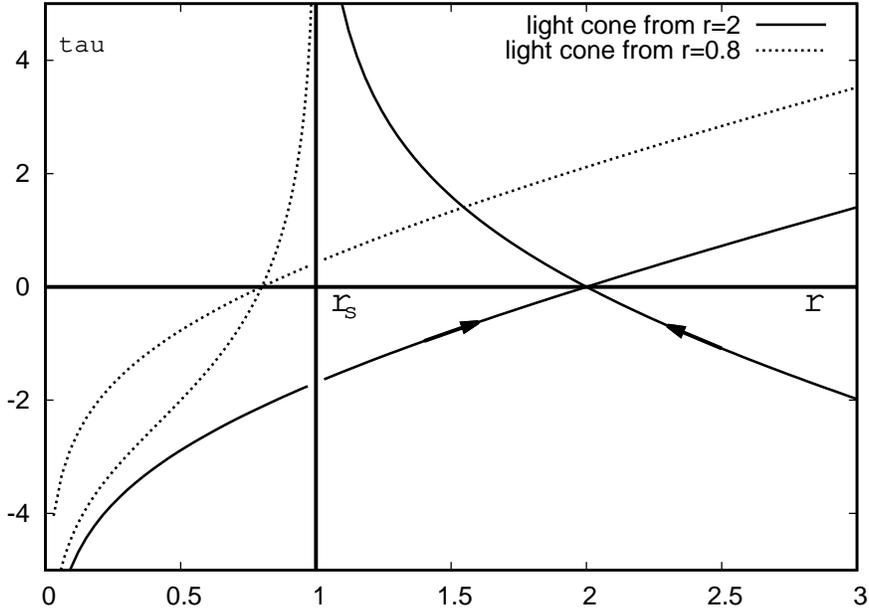}
  \caption{Light cones in the white hole coordinates $\tau,r$.  As
    before, the Schwarzchild radius is taken as unity.  One cone shown
    starts outside the hole at radius 2.  The second starts at radius
    0.8, inside the black hole.  Now out-going light waves do traverse
    the horizon.  In contrast incoming rays always asymptotically
    approach the horizon, which separates incoming light originating
    outside the hole from that on the inside.  As with the other
    coordinate choices, two way communication between the interior and
    exterior regions is forbidden.
    \medskip
  }
  \label{white}
\end{figure}

As this is just another choice of coordinates, all physical results
must be unchanged from what would see with either the Schwarzchild
coordinates or the times $w$ or $v_-$.  We have effectively reflected
Fig.~\ref{wlight} vertically about the $r$ axis.  Now a freely falling
object created inside the black hole can escape to reach points
outside.  This seems peculiar since we have only changed coordinates,
and this cannot not change any observations outside the horizon.  The
resolution is that the point where this object emerges from the
horizon maps into time being minus infinity in either $t$ or $w$.  In
essence, this object becomes an initial condition.  This peculiar
behavior is sketched in Fig.~\ref{white}.

 \vfill\eject


\hskip 1in
\section * {Summary}
Whichever time one selects, $t,\ w,\ v_-$ or $\tau$, is physically
arbitrary for an external observer.  All choices have space-like
constant time surfaces.  Two way communication is always forbidden
between the interior and exterior of the hole.



\vfill\eject


\end{document}